\documentclass{article}
\usepackage[utf8]{inputenc}
\usepackage{caption}
\usepackage{graphicx}
\usepackage{float} 
\usepackage{subcaption}
\usepackage{cite}
\usepackage[namelimits]{amsmath} %数学公式
\usepackage{amssymb}    
\usepackage{amsmath}
\title{Community Size and User Migration: Population Model Based on opinion Dynamics}

\author{FangYiKuang,Ding¹{ }{ }Yang,Li²\\
¹² Department of sociology\\
East China University of science and technology\\ Shanghai\\
China}
\date{June 2022}

\begin{document}

\maketitle

\begin{abstract}
Due to its significance in the recommendation system and community operations, user migration has garnered the interest of experts from numerous disciplines. However, contemporary research frequently overlooks the theory behind related prediction techniques, such as the Hidden Markov model. By combining the two fundamental processes of "opinion evolution" and "individual migration" in this research, the mechanistic explanation of online user migration is established and merged into a composite model. Simultaneously, some fundamental theorems and exploratory conclusions related to our model's consensus and steady population state are established via theoretical proof and numerical simulation. \\

\textbf{keywords:}\quad cyber population, user migration, opinion dynamics, bounded-confidence model, social simulation
\end{abstract}

\section{Introduction}
\subsection{Cyber Migration and its Model}
In recent years, due to the diversified growth of online communities and the emergence of the Web3.0 trend, relevant scholars have widely paid attention to cyber-demography \cite{2018Cyber}. However, two aspects exist in this field and need to be distinguished from each other: One is the study of the external population, which is about the overall growth and regional distribution of netizens \cite{2013Network}\cite{zhu}; The other is some internal issues of the cyber-population, which contains discussions pertinent to the scale of online communities and the flow of users in social media.  \cite{shen}  The content in our paper is closely related to the latter, and it focuses on the user migration in websites/online communities so that it can start a dialogue with many migration theories/models simultaneously.\\

As far as traditional demography is concerned, the theory describing migration can include multiple paradigms, including the "push-pull" structure emphasizing income factors, economic models focusing on "family decision-making," and multiple frameworks involving transnational social network/labor market segregation. Among them, the utility function can be introduced into the migration probability, making an effective analysis of residential isolation, and becoming an attempt in the field of analytical sociology \cite{2009The} . However, in the internal issues of the Internet population, due to the breakthrough of space limitation, the migration of individuals shifts from the movement of geographical location to the change of attention, and two special attributes emerge: (1) Users can stay in different communities/network areas at the same time (such as opening many webs and apps), but these areas are distributed with different attention. (2) The cost of cyber migration is greatly reduced, which makes the behavior almost dependent on individual wishes. \\

Based on the above characteristics, although the theory of cyber population is relatively lacking, the related prediction algorithms have been studied successfully. Such algorithms always emphasize two kinds of characteristics: one is the statistical characteristics of social networks (connectivity/degree distribution, etc.), and the other is the user behavior and its content production. The former enables the prediction model to acquire the mobile attributes of users, while the latter can establish matching among users in different communities (relying on similarity) and use real data for training and testing\cite{tian}. At the same time, this kind of algorithm can also be incorporated into the analysis strategies in classical demography, such as describing user migration as a "hidden Markov model" \cite{tian} , which is an application of stochastic process modeling.\\

On the whole, the features mentioned above have touched the core of the cyber population research: that is, to highlight an approach called "agent-based modeling" (ABM). In offline population problems, the dynamical system/spatial regression/grey prediction and other ways suitable for variables have once occupied the mainstream\cite{li}\cite{lai} \cite{dai}, which only discussed the number of population and its factors in some regions, while the cyber population research tried to describe the migration behavior and decision-making process of each individual, so as to serve the personalized user strategy. Therefore, the latter must develop a more microscopic theory in order to accommodate its own prediction and interpretation. In this regard, we should concentrate on the aspects of user migration, such as attention flow and individual preference, which are investigated by current opinion dynamics: By evaluating the interaction process of opinions, we may replicate the attitudes and relationships of individuals and apply our findings to a variety of social development challenges. 
\subsection{opinion Evolution and the Compound Process}
In order to introduce the evolutionary process of opinions into the theory of cyber migration, we need to select models from this field. As far as individual modeling is concerned, the game model/ising-sznajd model can be applied for the attitude formation of large-scale social networks\cite{2005Sznajd}\cite{2006Opinion}. However, this kind of framework has a class of unfavorable elements: because of its complex internal process, the current literature often abandons the mathematical proof of "consensus formation conditions" and turns to numerical simulation. It will leave a black box for readers, which is not conducive to the construction of the cyber population theory relying on mechanism explanation. Therefore, we can focus on a simplified paradigm called the "bounded confidence model," such as the H-K/Deffaut model\cite{2004Simulation} \cite{hegselmann2005opinion}, which enables individuals to exchange viewpoints with people whose viewpoints are adjacent to them and update individual views at the moment of t+1 through the weighted average of other people's views at the time of t. Its form is as follows:
$x_{i}(t+1)=\left(1 / N_{i}\right)\sum_{j}\cdot\left(x_{j}(t)\right), j \in N_{i}=\left\{j:\left|x_{j}(t)-x_{i}(t)\right|<d\right\}$\\

At the same time, the "bounded confidence model" mentioned above can be extended to more complex situations to fit reality. This includes many strategies such as "distinguishing private opinions from expressed opinions," "introducing random disturbances in the process of updating," and "adding stubborn individuals who have unchanged views"\cite{effect}\cite{stochastic}:the latter two are often included in the field of stochastic opinion dynamics. On this basis, some hidden problems can be discussed, like these two examples: (1) How to show the influence of individual i on individual j in the evolution of opinions? This will relate to the formation of social power (the mechanism of consent)\cite{2018Opinion}(2). If the opinions of neighboring individuals on the network are similar and closely related, how can this result be explained? This belongs to the field of "social fragmentation" and group cohesion\cite{sayama2022social}.\\

From this, we can find that the evolution of opinions, as the basic process of Internet society, can be extended to many classic issues in social theory. As far as population migration is concerned, it can capture the attitude of users to predict their direction of migration. Besides, it can effectively describe the control relationship between "factors" (such as network structure, opinion status) and "results" (population distribution/mobility, etc.) by means of mathematical deduction and computer simulation. Although there may be deviations in the absence of data calibration, it can provide effective evidence for establishing a theoretical explanation based on a compound mechanism \cite{2005Dissecting}. 
\section{Models and Theorems}
\subsection{opinion Model}

In order to introduce the random disturbance into our model and distinguish private opinions from expressed opinions, we combine H-K and Deffaut models to construct a model that not only has "the confidence threshold" (d in "$|x_j(t)-x_i(t)|$ < d"), but also updates by using opinions differences. Its basic framework is shown in the following figure. At the same time, we'll gradually establish the mathematical description of the
framework through some definitions and prove its\\

\begin{figure}[htbp]%调节图片位置，h：浮动；t：顶部；b:底部；p：当前位置
	\centering
	\includegraphics[width=0.9\linewidth]{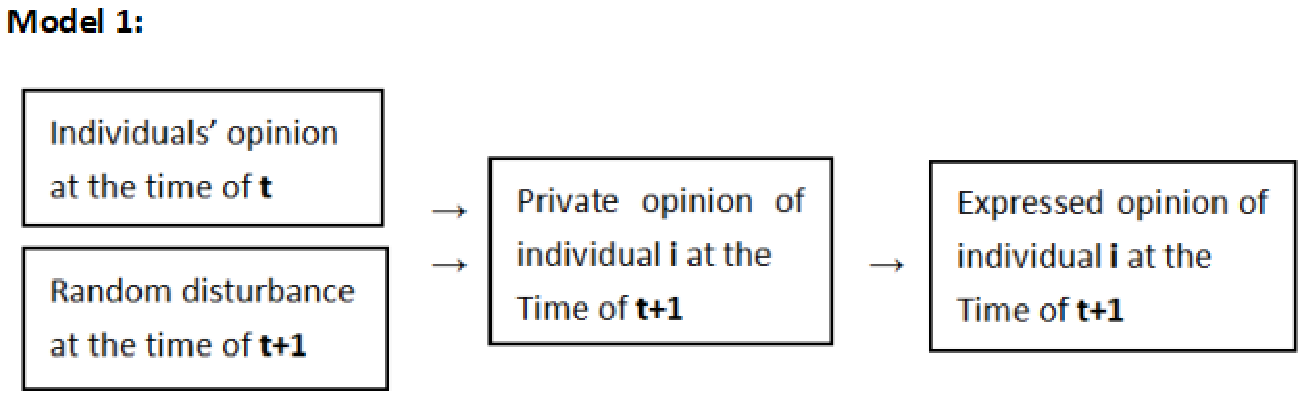}% 中括号中的为调节图片大小
	\label{chutian}%文中引用该图片代号
\end{figure}

\textbf{Definition 1}: G = (V, E) is a graph representing the social network and  $V=\{v_1, \ldots v_n\}$ is a vertex set representing individuals;$E=\{e_1, \ldots, e_m\}$ is an edge set $(\mathrm{e}=\{\mathrm{v_i}, \mathrm{v_j}\})$, which indicates that there is a social relationship between individuals. \\

\textbf{Definition 2}: $x_i$ (t) represents the private opinion of individual $v_i$ at the time of t and  $\hat{x}_{i}(t)$   represents the expressed opinion of individual $v_i$ at the time of t.\\

\textbf{Definition 3}:  
$\mathrm{N_i}=\left\{\mathrm{v_j}:\left|\hat{x}_{t}(t)-\hat{x}_{j}(t)\right|<\mathrm{d}\right\}$  is the neighborhood of vi under the opinion function.\\

\textbf{Definition 4}: If $\forall \varepsilon>0$,\quad $ \lim _{t \rightarrow \infty} P\left(d_{v}(t) \geq \varepsilon\right)=0$ ,\quad we say the opinion set $\left\{\hat{x}_{i}(t)\right.$ : $v_i \in $ $\left.\mathrm{V}\right\}$ exists a  quasi-consensus. 
Denote $d_{v}(t)=\sup _{i_{i} j \in V}\left|\hat{x}_{i}(t)-\hat{x}_{j}(t)\right|$. From the above four definitions, we can give the updated equation of the model, which is shown as follows (a basic form of Model 1). In addition, we'll present a basic theorem for the consensus formation of individual opinions in our model: \\

$\left\{\begin{array}{l}x_{i}(t+1)=\varphi \cdot\left[x_{i}(t)+\sum_{\mathrm{v} \in N_{i}} \frac{\hat{x}_{j}(t)-\hat{x}_{i}(t)}{\left|N_{i}\right|}\right]+(1-\varphi) \cdot \varepsilon(t) \\ \hat{x}_{i}(t+1)=\sigma \cdot x_{i}(t+1)\end{array}\right.$\\

Denote
$\varepsilon(t) \sim U\left(\inf _{i} x_{i}(t), \sup _{i} x_{\bar{i}}(t)\right)$\\

\textbf{Theorem 1 (consensus theorem)}: If $\varphi+\sigma<1$ , then the opinion set $\left\{\hat{x}_{i}(t)\right.$ : $v_i \in $ $\left.\mathrm{V}\right\}$ exists a quasi-consensus.\\

\textbf{Poof}:\\

(1) From the distribution of epsilon (t),\\

$\left|X_{j}(t+1)-X_{j}(t+1)\right|$

$\leq \varphi\left|x_{i}(t)-x_{j}(t)\right|+\varphi \cdot \sigma\left|\sum_{n \in N_{i}} \frac{x_{n}(t)}{\left|N_{t}\right|}-\sum_{m \in N_{j}} \frac{x_{m}(t)}{\left|N_{j}\right|}-x_{i}(t)+x_{j}(t)\right|+(1-\varphi)\left|\varepsilon_{i}(t)-\varepsilon_{j}(t)\right|$

$\leq \varphi\left|x_{i}(t)-x_{j}(t)\right|+\sigma \cdot \varphi\left(\sup _{i, j}\left|x_{i}(t)-x_{j}(t)\right|+\left|x_{i}(t)-x_{j}(t)\right|\right)+(1-\varphi) \sup _{i, j}\left|x_{i}(t)-x_{j}(t)\right|$

$\leq(\varphi+\varphi \cdot \sigma) \cdot\left|x_{i}(t)-x_{j}(t)\right|+\sigma \cdot \sup _{i, j}\left|x_{i}(t)-x_{j}(t)\right|$

$\leq(\varphi+\sigma+\varphi \cdot \sigma) \cdot \sup _{i, j}\left|x_{i}(t)-x_{j}(t)\right|$

$\leq(\varphi+\sigma)^{2} \cdot \sup _{i, j}\left|x_{i}(t)-x_{j}(t)\right|$\\

(2)Consider summarizing the conclusion below:\\

$\left|x_{i}(t)-x_{j}(t)\right| \leq(\varphi+\sigma)^{2 t} \cdot k, k=\sup _{i, j}\left|x_{i}(0)-x_{j}(0)\right|$\\

When n=1 it is known by (1)\\

$\left|x_{i}(1)-x_{j}(1)\right| \leq(\varphi+\sigma)^{2} \cdot \sup _{i, j}\left|x_{i}(0)-x_{j}(0)\right|=(\varphi+\sigma)^{2} \cdot k$\\

Assuming n=t-1, if the conclusion holds, then there are\\

$\left|x_{i}(t-1)-x_{j}(t-1)\right| \leq(\varphi+\sigma)^{2 t-2} \cdot k$\\

Then\\

$\left|x_{i}(t)-x_{j}(t)\right| \leq(\varphi+\sigma)^{2} \cdot \sup _{i, j}\left|x_{i}(t-1)-x_{j}(t-1)\right| \leq(\varphi+\sigma)^{2 t} \cdot k$\\

(3) Combining the model with the conclusion in (2), the following results can be obtained: \\

$E\left|\hat{x}_{i}(t)-\hat{x}_{j}(t)\right| \leq \sigma \cdot(\varphi+\sigma)^{2 t} \cdot k$\\

Furthermore, according to the Kolmogorov inequality, \\

$P\{d v(t) \geq \varepsilon\}=P\left\{\sup _{i, j} \left|\hat{x}_{i}(t)-\hat{x}_{j}(t)\right| \geq \varepsilon\right\}$\\

$=P\left(\max _{i, j}\left|\hat{x}_{i}(t)-\hat{x}_{j}(t)\right| \geq \varepsilon\right) \leq \frac{\sigma \cdot(\varphi+\phi)^{2 t} \cdot k}{\varepsilon}$\\

then\\

$\varphi+\sigma<1$\\

hence\\

$\lim _{t \rightarrow \infty} p\left\{d_{v}(t) \geq \varepsilon\right\}=0$\\

which shows the quasi-consensus, Q.E.D\\

Through the proof of the above theorem, we can obtain a sufficient condition for the existence of quasi-consensus, which can be effectively applied in the analysis of the following model. 
\subsection{Migration Model: a Compound attempt }

After establishing a dynamic model to describe the evolution process of opinions, we can further describe the state of users when making migration decisions. Among them, the model of "population segregation" in analytical sociology deserves attention, It can be normalized with softmax by a utility function, and then the form of individual migration probability is given. Considering the particularity of online mobility, we refer to this framework and include two elements: "social distance" and "opinion distance. "The former describes "accessibility" (the "possibility of contact") between individual i and group j, while the latter measures the opinion differences between them (which can reflect the preference degree of i for j). Therefore, we built a compound model to describe the demographic changes of two "mutually competitive communities," which is as follows.\\

\begin{figure}[htbp]%调节图片位置，h：浮动；t：顶部；b:底部；p：当前位置
	\centering
	\includegraphics[width=0.9\linewidth]{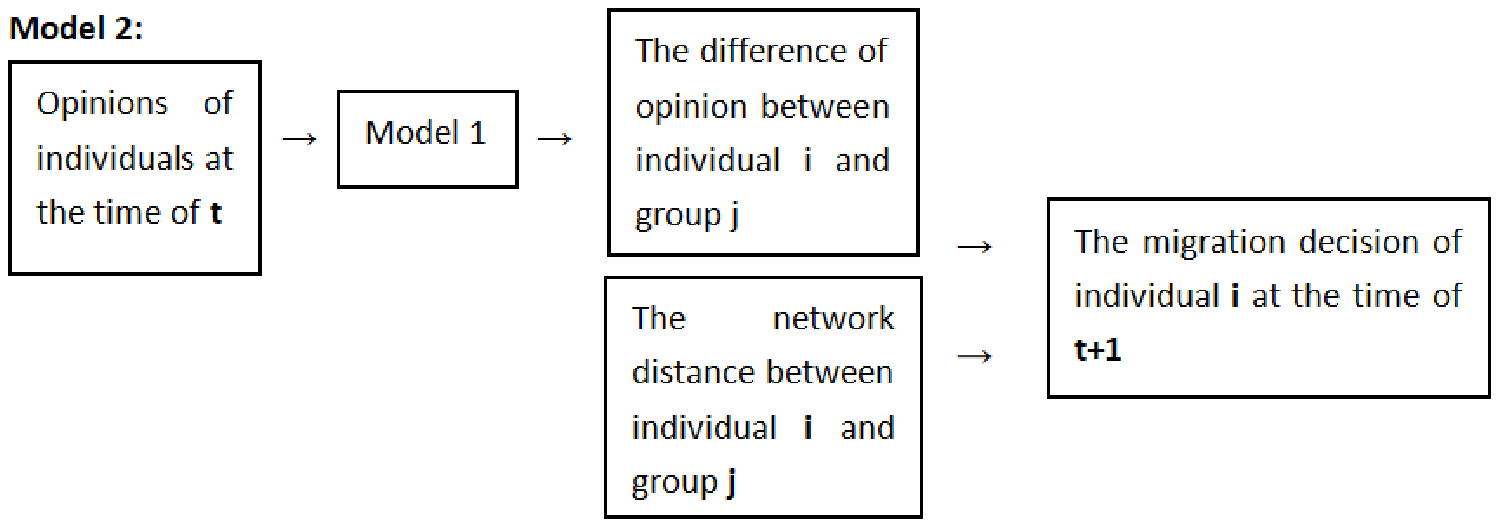}% 中括号中的为调节图片大小
	\label{chutian}%文中引用该图片代号
\end{figure}

\textbf{Definition 5}: \\

$Q_{j}=\left\{v_{1} ; \cdots, v_{m}\right\} \subset V$  is a community \\

$Y_{j}=\left|Q_{j}\right|$is the population of community j.\\

$x_{i . j}=\left\{\begin{array}{l}1, v_{j} \,\text { migrates \, into } Q_j \\ 0, \text { other \, conditions }\end{array}, d\left(v_{i}, Q_{j}\right)=\frac{\sum_{v_{j} \in Q_{j}} d\left(v_{i}, v_{j}\right)}{\left|Q_{j}\right|}, \operatorname{avg}\left(\hat{x}_{j}\right)=\frac{\sum_{j \in Q_{j}} \hat{x}_{j}}{\left|Q_{j}\right|}\right.$\\

\textbf{Definition 6}: G is a star graph if and only if there is a unique node $v_i$ called a center, and for a node $v_j$ that is not equal to vi, there is an edge between $v_j$ and $v_i$, but no edge between any $v_j$. 

According to the above definition, we can give a probability expression to describe the user's migration choice (the two competitive groups are $Q_j$ and $Q_k$, and $v_1$ is the center), which includes the following points: 

\textbf{(1) Universal Pattern:}\\

$P\left(X_{i k}=1 \mid X_{1 k}=1\right)=\frac{\exp \left[\delta_{1} \cdot d\left(v_{i}, Q_{j}\right)+(1-\delta) \cdot\left|\hat{x}_{i}-\operatorname{avg}\left(\hat{x}_{j}\right)\right|\right]}{\sum_{j=1}^{2} \exp \left[\delta \cdot d\left(v_{i}, Q_{j}\right)+(1-\delta) \cdot\left|\hat{x}_{i}-\operatorname{avg}\left(\hat{x}_{j}\right)\right|\right]}$\\

\textbf{(2) Special Case:} In the star map,  $x_{i j}$ it is mutually independent under the condition $x_{1 j}$ ; this is because any $v_j$ in the figure is only connected to the center $v_1$, so it is independent of each other.
Therefore, under the restriction of special conditions, we can deduce a conclusion called the "leader theorem," which describes the migration behavior in a star graph.\\

\textbf{Theorem 2('leader theorem'):}  If the social network is a star graph and there are only two communities, $Q_j$ and $Q_k$, and any individual $v_i$ and $v_j$ update their opinions according to model 1, when t is sufficiently large (that is, "there is a positive integer N, t > N"), when the center $v_1$ enters the community $Q_k$, the expectation of the population of the community $Q_k$ is\\

$E Y_{k}=\sum_{z=1}^{n}\left(\frac{e^{2 \delta}}{M}\right)^{z-1} \cdot\left(1-\frac{e^{2 \delta}}{M}\right)^{n-z} \cdot z$ , \\

where \\

M = $\sum_{j=1}^{2} \exp \left[\delta \cdot d\left(x_{i}, Q_{j}\right)+(1-\delta) \cdot\left|\hat{x}_{i}-\operatorname{avg}\left(\hat{x}_{j}\right)\right|\right]$ .\\

\textbf{Proof}: 
According to the 'consensus theorem,' when t is sufficiently large, the absolute value of the difference between opinions tends to 0; And according to the properties of the star graph, we know d ($v_i$, $v_j$) = 2 for $v_i$ and $v_j$ of any non-central node. Therefore, the following formulas can be obtained: \\

Based on $P\left(Y_{k}=z\right)=P\left(X_{1 k}=1\right) \cdot \prod_{i=2}^{2} P\left(x_{i k}=1 \mid x_{1 k}=1\right) \prod_{i=z+1}^{n} P\left(x_{i k}=0 \mid x_{1 k}=1\right)$\\

we know $P\left(X_{i k}=1 / X_{1 k}=1\right)=\frac{\exp (2 \delta)}{M_{i}}=\frac{\exp (2 \delta)}{M}$\\

From\quad $P\left(X_{1 k}=1\right)=1$,\\

$P\left(Y_{k}=z\right)=\prod_{i=2}^{z} P\left(x_{i k}=1 \mid x_{1 k}=1\right)\cdot  \prod_{i=z+1}^{n} P\left(x_{i k}=0 \mid x_{1 k}=1\right)=\left(\frac{e^{2 \delta}}{M}\right)^{z-1} \cdot\left(1-\frac{e^{2 \delta}}{M}\right)^{n-z}$holds\\

According to the definition of expectation, we know \\

$E Y_{k}=\sum_{z=1}^{n}\left(\frac{e^{2 \delta}}{M}\right)^{z-1} \cdot\left(1-\frac{e^{2 \delta}}{M}\right)^{n-z} \cdot z$, Q.E.D.\\

Using the above theorem, we can apply the compound model to the population prediction of cyber migration in a special mode (for example, the vertical strategy of Xiaohongshu has a star structure). Due to the complexity of the problem, a more common condition of our model will be discussed through numerical experiments later, which can more clearly show the nonlinear influence of ‘social/opinion distance’ on individual behavior. 
\section{Simulation and Experimental Design}
\subsection{Parameter Settings}
The parameters in our paper are divided into two categories: one is structural parameters, which describe the features of the social network and its internal regions (namely communities); The second is individual parameters, which show the characteristics of individuals when integrating other people's opinions and carrying out migration behavior. These two types of parameters are represented by Greek letters in the above model, which is convenient for identification.\\

As far as structural parameters are concerned, we first set the type of graph G, which is a small-world network: to enhance the representativeness of our simulation, the experiment will include two cases, which respectively contain 50/500 nodes(to represent different population sizes), and the probability of random reconnection is 0.3. In addition, the simulated user migration behavior is limited to "duopoly competition mode," which means 'the number of communities is 2'. Finally, we set the noise/disturbance in the evolution of opinions, which can be randomly selected between the minimum and maximum values of opinions at the time of t and accords with the "uniform distribution" mentioned above. \\

In terms of individual parameters, we set different confidence thresholds d in model 1, including 1, 0.8, and 0.3, where 1 is an extreme case, indicating that individuals in the network can influence each other's opinions at any time. At the same time, we set parameters $\varphi$ and $\sigma$: the former includes 1, 0.5, 0.4, 0.09, etc. (1-$\varphi$ represents the degree of individual being affected by noise), while the latter has values of 1, 0.9, 0.4, etc. (representing the reduction rate from private opinions to expressed opinions, which can be regarded as the effect of "the spiral of silence"). Finally, we consider the parameters $\delta$ when individuals make migration decisions, and $\delta$/1-$\delta$ represent the influence of social distance/opinion distance, respectively. Therefore, we set them to 0.8 and 0.3, representing individuals with the intimacy tendency and opinion convergence tendency in current social media. \\
\subsection{Measurement and Visualization}
In the aspect of presenting simulation results, we use two methods: one is the measurement of core indices, and the other is the data visualization of experimental results. In terms of indices, we calculate the ‘net growth rate of population’(NGR) and ‘net migration rate’(NMR) of multiple population experiments at each time, and the computing formulas are : \\

(1) $N G R=$ (Popu $(t+1)-$ Popu $(t)) / 0.5^{*}[$ Popu $(t+1)+$ Popu $(t)]$\\

(2) NMR $=$ (immi $(t)-e m m i(t)) / \operatorname{Popu}(t)=\operatorname{Popu}(t+1)-\operatorname{Popu}(\mathrm{t}) / \mathrm{Popu}(\mathrm{t})$\\

Among them, Popu (t), immi (t), and emmi (t) respectively represent the total population, immigration population, and emigration population at the time of t, and the model in this paper does not consider natural growth. For data visualization, we will use composite line charts to show the changes in individual opinions and population numbers on the horizontal axis t. At the same time, each individual and group are presented in different colors, which is convenient for visual distinction. 
\section{Results and Conclusions}
\subsection{Opinion Simulation }
\subsubsection{Minority consensus and its phase transition (n=50)}
In a small network of 50 people, we first simulate the results obtained in the above mathematical analysis, and it can be seen in the following figure. Although individual opinions fluctuate greatly at the initial stage, they can reach the "consensus" state in a short period of time (after updating 20 times), which verifies the reliability of the theorem proved above. Besides, we control the weight of random disturbance terms and find that the model has a high convergence rate when it is large. This result coincides with the daily public opinion: disturbance term can be regarded as a special intervention (such as purposeful speech from an official bot). At the same time, we also find that the opinion reduction rate $\sigma$ greatly influences consensus formation: when $\sigma$ = 0.4 (small), the model only needs 5 iterations to converge, which actually simulates the evolution process of opinions under the strict scrutiny.\\ 

\begin{figure}[htbp]
	\centering
	\begin{subfigure}{0.32\linewidth}
		\centering
		\includegraphics[width=0.9\linewidth]{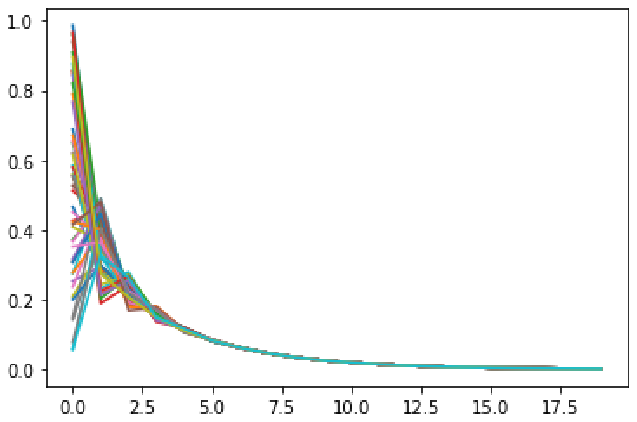}
		\caption{d = 1,$\varphi$ = 0.5,$\sigma$ = 0.4}
		\label{chutian3}%文中引用该图片代号
	\end{subfigure}
	\centering
	\begin{subfigure}{0.32\linewidth}
		\centering
		\includegraphics[width=0.9\linewidth]{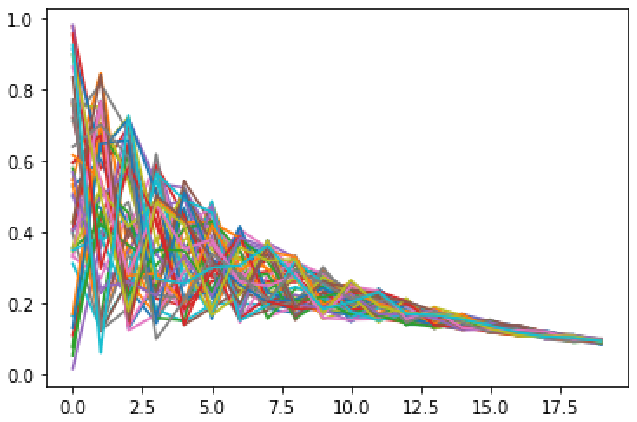}
		\caption{d = 1,$\varphi$ = 0.4,$\sigma$ = 0.5}
		\label{chutian3}%文中引用该图片代号
	\end{subfigure}
	\centering
	\begin{subfigure}{0.32\linewidth}
		\centering
		\includegraphics[width=0.9\linewidth]{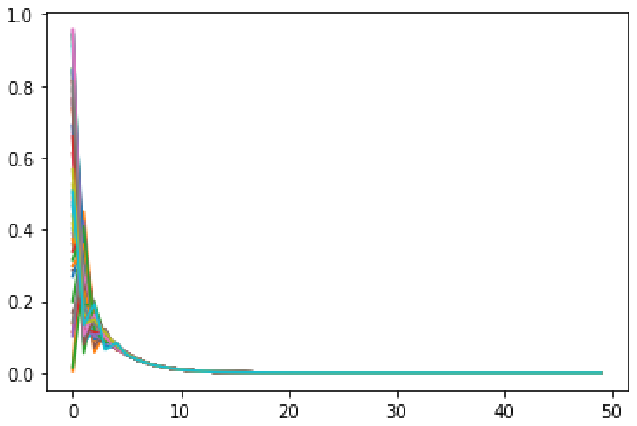}
		\caption{d = 0.8,$\varphi$ = 0.09,$\sigma$ = 0.9}
		\label{chutian3}%文中引用该图片代号
	\end{subfigure}
	\caption{}
	\label{da_chutian}
\end{figure}

After confirming the reliability of the previous theorem, we can carry out an ‘exploratory simulation’ to observe the consensus state when the condition ($\phi+\sigma<1$) of our theorem is not met. First of all, we simulate under the condition that$\phi+\sigma=1.4$ and d = 1: the condition is also strict, which requires all individuals to exchange opinions with each other (though it can be achieved when only 50 people stay in the system). According to the visualization results, we find that the opinions will still reach a consensus, and the convergence value is at a low level due to the disturbance of random attributes (0-1 can express the strength of a certain emotion or attitude, such as the preference degree to a person). Therefore, we simulate the situation without random disturbance ($\varphi$ = 1) and find that the consensus level/convergence value increases, but there is a continuous fluctuation, which may prolong the time of consensus formation. This further verifies the conclusion of the previous conclusion (disturbance term is beneficial to consensus formation), and we find another effective property: the uniform random disturbance can inhibit individuals' opinions in the network and make it go to a lower level. \\

\begin{figure}[htbp]
	\centering
	\begin{subfigure}{0.325\linewidth}
		\centering
		\includegraphics[width=0.9\linewidth]{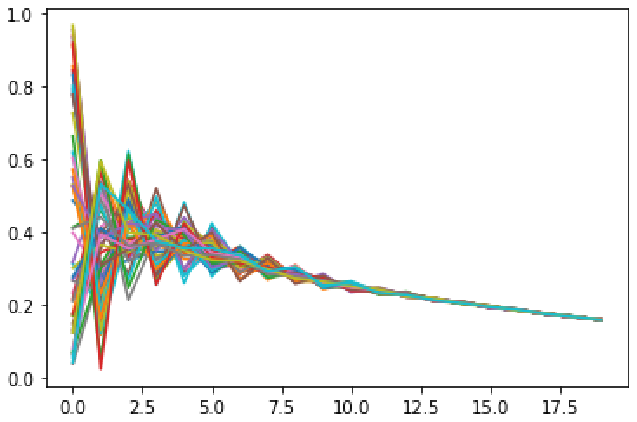}
		\caption{d = 1, $\varphi$ = 0.5, $\sigma$ = 0.9}
		\label{chutian3}%文中引用该图片代号
	\end{subfigure}
	\centering
	\begin{subfigure}{0.325\linewidth}
		\centering
		\includegraphics[width=0.9\linewidth]{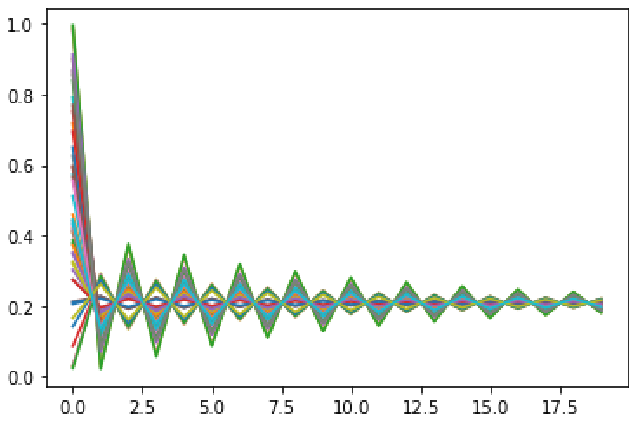}
		\caption{d = 0.3, $\varphi$ = 1, $\sigma$ = 0.9}
		\label{chutian3}%文中引用该图片代号
	\end{subfigure}
	\centering
	\begin{subfigure}{0.325\linewidth}
		\centering
		\includegraphics[width=0.9\linewidth]{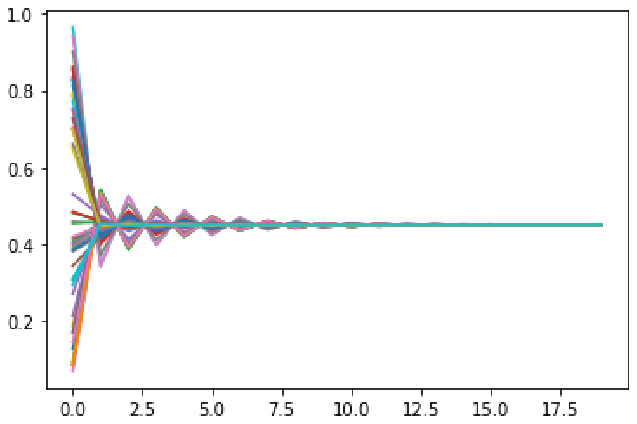}
		\caption{d = 0.3, $\varphi$ = 1, $\sigma$ = 0.8}
		\label{chutian3}%文中引用该图片代号
	\end{subfigure}
	\caption{}
	\label{da_chutian}
\end{figure}

According to the above analysis, we find the generality of "consensus formation" in our framework, so it is very important to explore the model conditions of "no consensus" for fitting reality (although it is difficult in mathematical deduction). In this regard, we design the following comparative experiments: (1) through a set of extreme parameters, we get the non-convergent model results; (2)Then, we adjust several parameters separately to observe the sensitivity of model results to the changes of each parameter; (3) By comparing the conditions and results in (2), the core factors leading to the result of "consensus does not exist" can be obtained. After selecting the results, we present them as follows: \\

\begin{figure}[htbp]
	\centering
	\begin{subfigure}{0.325\linewidth}
		\centering
		\includegraphics[width=0.9\linewidth]{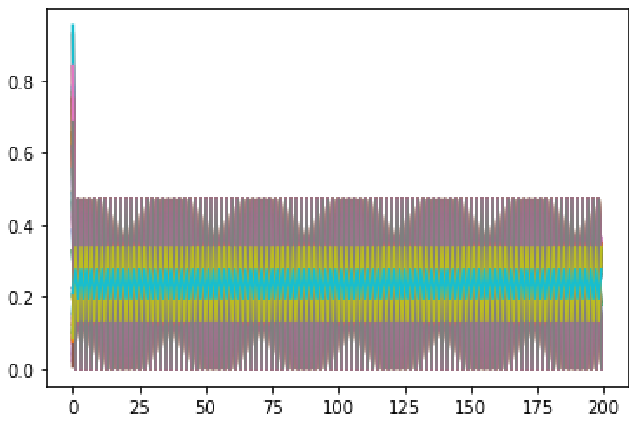}
		\caption{d = 1, $\varphi$ = 1, $\sigma$ = 1}
		\label{chutian3}%文中引用该图片代号
	\end{subfigure}
	\centering
	\begin{subfigure}{0.325\linewidth}
		\centering
		\includegraphics[width=0.9\linewidth]{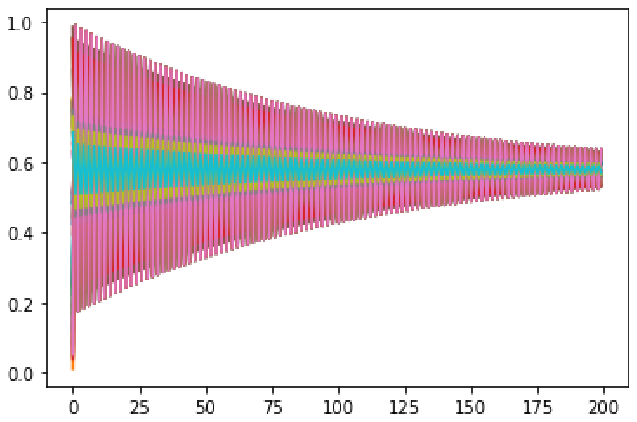}
		\caption{$\sigma$ = 0.99, $\varphi$ = d = 1}
		\label{chutian3}%文中引用该图片代号
	\end{subfigure}
	\centering
	\begin{subfigure}{0.325\linewidth}
		\centering
		\includegraphics[width=0.9\linewidth]{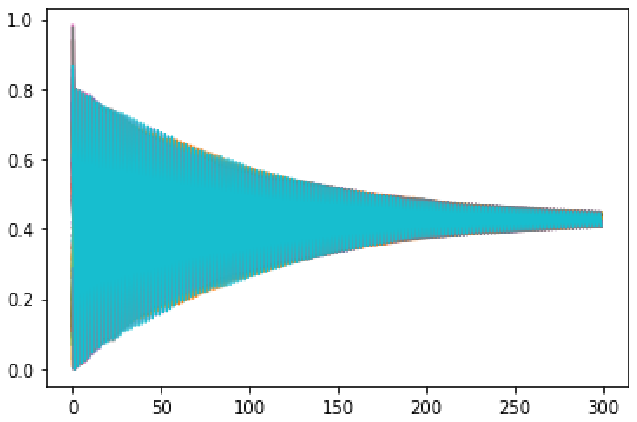}
		\caption{$\varphi$ = 0.99, d = $\sigma$ = 1}
		\label{chutian3}%文中引用该图片代号
	\end{subfigure}
	\label{da_chutian}
	\centering
	\begin{subfigure}{0.325\linewidth}
		\centering
		\includegraphics[width=0.9\linewidth]{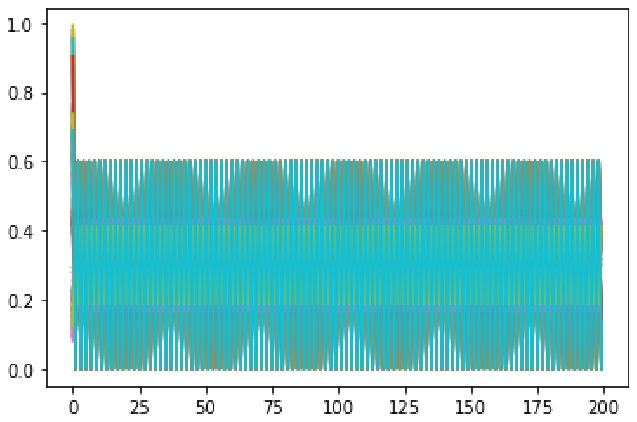}
		\caption{d=0. 99,$\varphi$ = $\sigma$ = 1}
		\label{chutian3}%文中引用该图片代号
	\end{subfigure}
	\centering
	\begin{subfigure}{0.325\linewidth}
		\centering
		\includegraphics[width=0.9\linewidth]{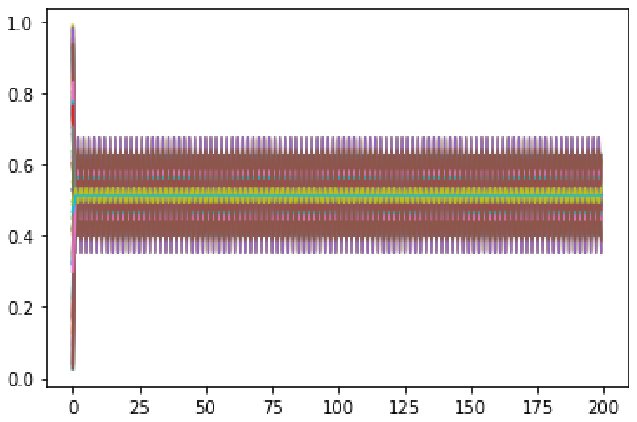}
		\caption{d=0. 5,$\varphi$ = $\sigma$ = 1}
		\label{chutian3}%文中引用该图片代号
	\end{subfigure}
	\centering
	\begin{subfigure}{0.325\linewidth}
		\centering
		\includegraphics[width=0.9\linewidth]{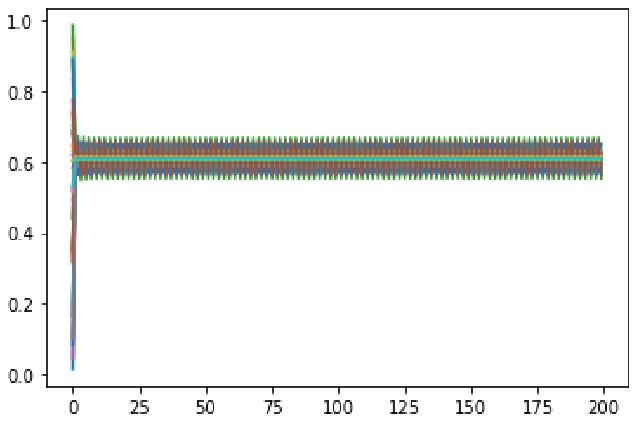}
		\caption{d=0. 1,$\varphi$ = $\sigma$ = 1}
		\label{chutian3}%文中引用该图片代号
	\end{subfigure}
	\caption{}
	\label{da_chutian}
\end{figure}

As can be seen from the above figure, we first set all three parameters equal to 1, which means (1) "any individual can exchange views," (2) "there is no random disturbance in the system," (3) "expressed opinions are same to private opinions ."We find that the opinions will fluctuate disorderly within a specific range, which is similar to the distribution of expressed opinion in public space under uncontrolled conditions: this can be essentially seen as a special "phase change ."Then, under the condition of keeping d unchanged, we make $\sigma$ and $\varphi$ equal to 0.99, respectively (the other term remains unchanged), and find that the opinions can gradually evolve into consensus (although it is extremely slow, exceeding 200 moments). It shows that the values of $\sigma$ and $\varphi$ are critical and have a high contribution to "non-consensus"; At the same time, when $\varphi$ and $\sigma$ are equal to 1 (that is, unchanged), we gradually reduce the value of d to 0.1 and find that it cannot reach a consensus/convergence state, but it can reduce the fluctuation range of opinions. Generally speaking, $\varphi$ (undisturbed degree) and $\sigma$ (reduction rate) jointly determine whether the evolution of opinions has abrupt changes (from consensus to non-consensus), while d (confidence threshold) determines the range of fluctuations of opinions under the condition of abrupt changes. \\

In the real world, the above-mentioned "phase transition" describes the "disorderly evolution" in the dynamic process of opinions, and there is no attractor (which can be understood as a common trend in reality). This means that when the expression of opinions is neither silent nor reduced due to external pressure nor randomly disturbed by "neutral people" (such as Xijin Hu in Chinese social media), it will be an extremely difficult process to reach a "consensus," even if there is a lot of exchange of information and opinions in the process. At the same time, the higher the confidence threshold of an individual (which means that the opinions that influence him are wider), the greater the fluctuation range of the overall opinions. These two results may directly show two challenges faced by "open society" and "pluralism": (1) Complex views make consensus far away from us(although it may not meet polarization); (2) The promotion of "tolerance" in liberalism (the parameter d in our model) may not change the trend shown above.
\subsubsection{Simulation of large-scale networks (n=500) }
In the range described by the theorem ($\varphi$ + $\sigma$ < 1), the consensus of opinions is not affected by the number of individuals in the network, so we only simulate the parameters that do not meet the theorem conditions. At the same time, in order to ensure the control characteristics of the experiment, we make the parameters consistent with some cases above(n=50) so as to make a comparison. According to the following results, we find that there is no obvious difference between its trend and the previous results, which shows that the model has no strong sensitivity to the growth of nodes. Therefore, the consensus under generalization conditions ($\phi<1$ or $\sigma<1$) may be a broad conclusion, which is expected to be proved mathematically in the future. \\

\begin{figure}[htbp]
	\centering
	\begin{subfigure}{0.325\linewidth}
		\centering
		\includegraphics[width=0.9\linewidth]{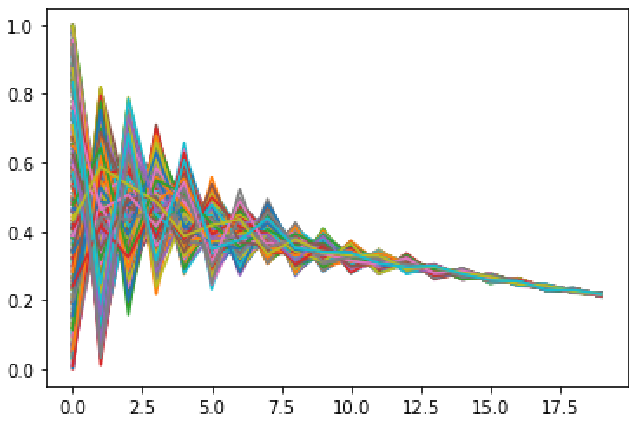}
		\caption{d = 1,$\varphi$ = 0.5,$\sigma$ = 0.9}
		\label{chutian3}%文中引用该图片代号
	\end{subfigure}
	\centering
	\begin{subfigure}{0.325\linewidth}
		\centering
		\includegraphics[width=0.9\linewidth]{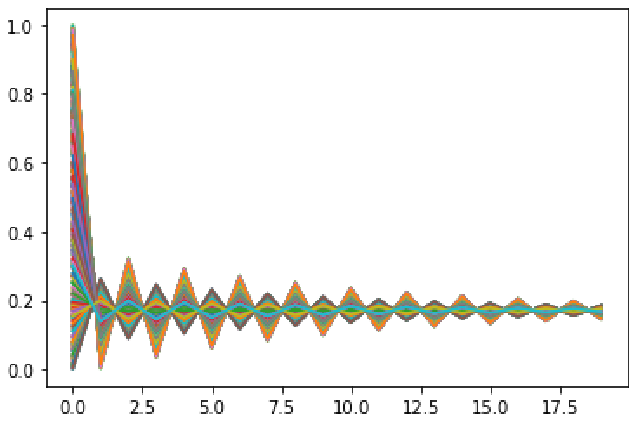}
		\caption{d = 0.3,$\varphi$ = 1, $\sigma$ = 0.9}
		\label{chutian3}%文中引用该图片代号
	\end{subfigure}
	\centering
	\begin{subfigure}{0.325\linewidth}
		\centering
		\includegraphics[width=0.9\linewidth]{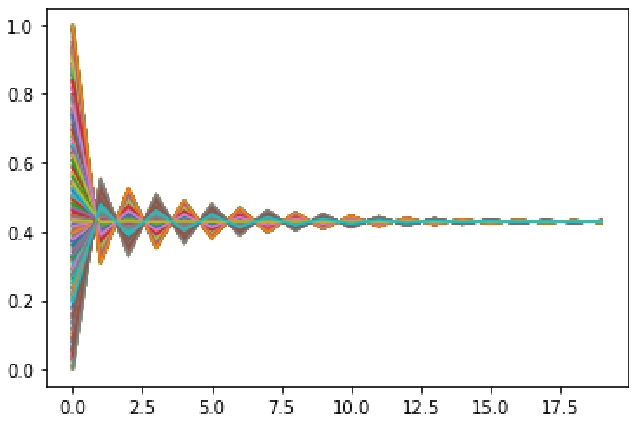}
		\caption{d = 0.3,$\varphi$ = 1,$\sigma$ = 0.8}
		\label{chutian3}%文中引用该图片代号
	\end{subfigure}
	\caption{}
	\label{da_chutian}
\end{figure}

\subsection{Population Simulation}

\subsubsection{Consensus and Migration(n=50)}
In the population model built in the second part, the probability of individual i migrating to community j depends on the "social distance" (the distance on the graph) and "opinion distance" (the difference between opinion values). For a long time, if the individuals in the network can reach a consensus, this "opinion distance" will tend to 0: it is a basic conclusion in our "ideal model ."At the same time, it provides us with an important feature: for any two composite models (population + opinion), no matter what the parameters of their opinion model are, as long as they converge at the time of t, the evolution differences between the two population models is small after the time of t. Therefore, we only need to select a set of representative opinion parameters ($\phi, \sigma, d$). In the following simulation, the following values will be satisfied: $d=1, \phi=0.5, \quad \sigma=0.9$. \\

Then, we simulated different $\delta$ values to observe the migration on the small-world network. At the same time, we can also calculate the net growth rate of population(NGR) and net migration rate(NMR) of the two groups every P times (taking P=5), so as to observe their stability. Results are presented as follows:\\

\begin{figure}[htbp]
	\centering
	\begin{subfigure}{0.45\linewidth}
		\centering
		\includegraphics[width=0.9\linewidth]{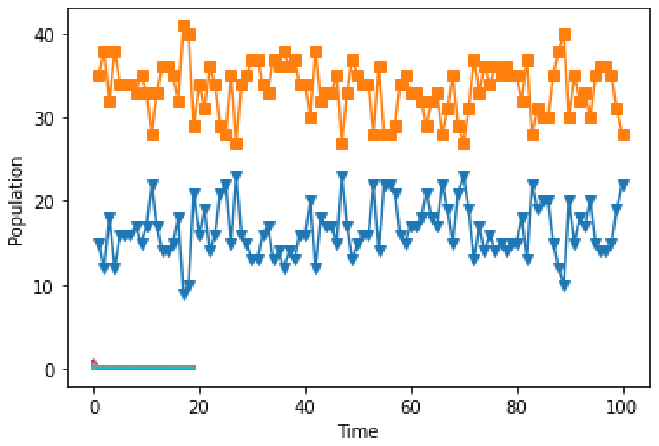}
		\caption{N = 50, $\delta$= 0.3,  d=1 ,$\varphi$= 0.5 ,$\sigma$ = 0.9}
		\label{chutian3}%文中引用该图片代号
	\end{subfigure}
	\centering
	\begin{subfigure}{0.45\linewidth}
		\centering
		\includegraphics[width=0.9\linewidth]{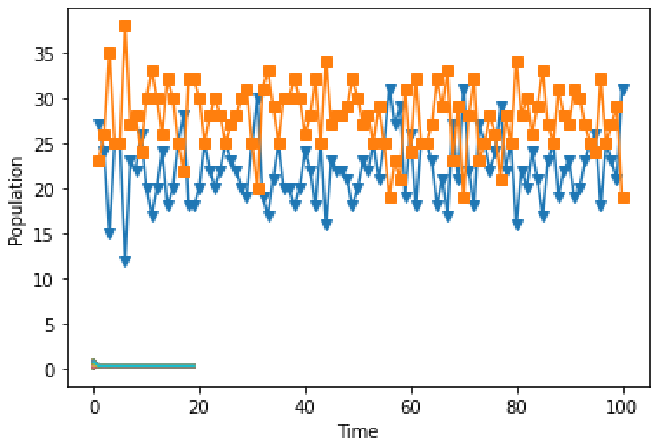}
		\caption{N = 50,$\delta$= 0.8,  d=1 ,$\varphi$ = 0.5 ,$\sigma$ = 0.9}
		\label{chutian3}%文中引用该图片代号
	\end{subfigure}
	\caption{}
	\label{da_chutian}
\end{figure}
\begin{table}[]
\begin{tabular}{|ll|ll|ll|ll|}
\hline
\multicolumn{2}{|l|}{NGR (0.3)}         & \multicolumn{2}{l|}{NMR (0.3)}         & \multicolumn{2}{l|}{NGR (0.8)}         & \multicolumn{2}{l|}{NMR (0.8)}         \\ \hline
\multicolumn{1}{|l|}{Group A} & Group B & \multicolumn{1}{l|}{Group A} & Group B & \multicolumn{1}{l|}{Group A} & Group B & \multicolumn{1}{l|}{Group A} & Group B \\ \hline
\multicolumn{1}{|l|}{-0.22}   & 0.08    & \multicolumn{1}{l|}{-0.2}    & 0.08    & \multicolumn{1}{l|}{-0.11}   & 0.12    & \multicolumn{1}{l|}{-0.11}   & 0.13    \\ \hline
\multicolumn{1}{|l|}{0.06}    & -0.02   & \multicolumn{1}{l|}{0.06}    & -0.02   & \multicolumn{1}{l|}{-0.04}   & 0.03    & \multicolumn{1}{l|}{-0.04}   & 0.03    \\ \hline
\multicolumn{1}{|l|}{0}       & 0       & \multicolumn{1}{l|}{0}       & 0       & \multicolumn{1}{l|}{-0.28}   & 0.20    & \multicolumn{1}{l|}{-0.25}   & 0.23    \\ \hline
\multicolumn{1}{|l|}{-0.27}   & 0.15    & \multicolumn{1}{l|}{-0.23}   & 0.17    & \multicolumn{1}{l|}{0.11}    & -0.06   & \multicolumn{1}{l|}{0.11}    & -0.06   \\ \hline
\multicolumn{1}{|l|}{-0.37}   & 0.22    & \multicolumn{1}{l|}{-0.31}   & 0.25    & \multicolumn{1}{l|}{-0.08}   & 0.07    & \multicolumn{1}{l|}{-0.08}   & 0.08    \\ \hline
\multicolumn{1}{|l|}{0.20}    & -0.08   & \multicolumn{1}{l|}{0.23}    & -0.08   & \multicolumn{1}{l|}{-0.44}   & 0.43    & \multicolumn{1}{l|}{-0.36}   & 0.55    \\ \hline
\multicolumn{1}{|l|}{-0.07}   & 0.02    & \multicolumn{1}{l|}{-0.07}   & -0.03   & \multicolumn{1}{l|}{-0.10}   & 0.06    & \multicolumn{1}{l|}{-0.1}    & 0.06    \\ \hline
\multicolumn{1}{|l|}{-0.06}   & 0.03    & \multicolumn{1}{l|}{-0.05}   & 0.03    & \multicolumn{1}{l|}{-0.43}   & 0.30    & \multicolumn{1}{l|}{-0.36}   & 0.36    \\ \hline
\multicolumn{1}{|l|}{0.14}    & -0.05   & \multicolumn{1}{l|}{0.15}    & -0.05   & \multicolumn{1}{l|}{0.1}     & -0.06   & \multicolumn{1}{l|}{0.11}    & -0.06   \\ \hline
\multicolumn{1}{|l|}{0}       & 0       & \multicolumn{1}{l|}{0}       & 0       & \multicolumn{1}{l|}{0.21}    & -0.27   & \multicolumn{1}{l|}{0.24}    & -0.24   \\ \hline
\multicolumn{1}{|l|}{0.06}    & -0.03   & \multicolumn{1}{l|}{0.05}    & -0.03   & \multicolumn{1}{l|}{0.32}    & -0.24   & \multicolumn{1}{l|}{0.38}    & -0.21   \\ \hline
\multicolumn{1}{|l|}{-0.23}   & 0.12    & \multicolumn{1}{l|}{-0.21}   & 0.12    & \multicolumn{1}{l|}{0.45}    & -0.35   & \multicolumn{1}{l|}{0.58}    & -0.30   \\ \hline
\multicolumn{1}{|l|}{-0.19}   & 0.08    & \multicolumn{1}{l|}{-0.17}   & 0.09    & \multicolumn{1}{l|}{-0.07}   & 0.08    & \multicolumn{1}{l|}{-0.07}   & 0.08    \\ \hline
\multicolumn{1}{|l|}{0}       & 0       & \multicolumn{1}{l|}{0}       & 0       & \multicolumn{1}{l|}{-0.43}   & 0.30    & \multicolumn{1}{l|}{-0.36}   & 0.36    \\ \hline
\multicolumn{1}{|l|}{0}       & 0       & \multicolumn{1}{l|}{0}       & 0       & \multicolumn{1}{l|}{0.3}     & -0.2    & \multicolumn{1}{l|}{0.35}    & -0.18   \\ \hline
\end{tabular}
\end{table}
According to the above results, we can see that: when $\delta$ is small (It means social distance has little influence and opinion distance has a great influence on the decision of online migration), although the population of the community will not converge to a certain value, its mobility is small as a whole. It can be seen from the yellow curve that the scales of groups A and B are relatively stable: this shows that in a model existing the consensus if individuals consider opinion differences more to choosing communities, their mobility probability is low, which can explain the stability of users in BBS Module and other similar areas. However, when the $\delta$ is large, due to some common characteristics of the small-world network (the distance between nodes is small), individuals will migrate based on the social distance, which is difficult for them to form steady-state contact with specific groups, resulting in strong population distribution fluctuations and high mobility. \\

At the same time, in order to study the effect of network structure (that is, the social distance between individuals) on the migration, we also consider an extreme case that is "the population evolution of two competing groups without the influence of social distance ($
\delta$=0)", and the result is as follows. Based on this, we find that in the small-world network, the population among groups is still changing continuously, and there exists an intersection (i.e., $n_1$=$n_2$=25). It shows that the network structure (i.e., social relationship) is a core factor affecting migration, while the opinion differences affect the scope of population change between the two groups. \\

\begin{figure}[htbp]
	\centering
	\includegraphics[width=0.9\linewidth]{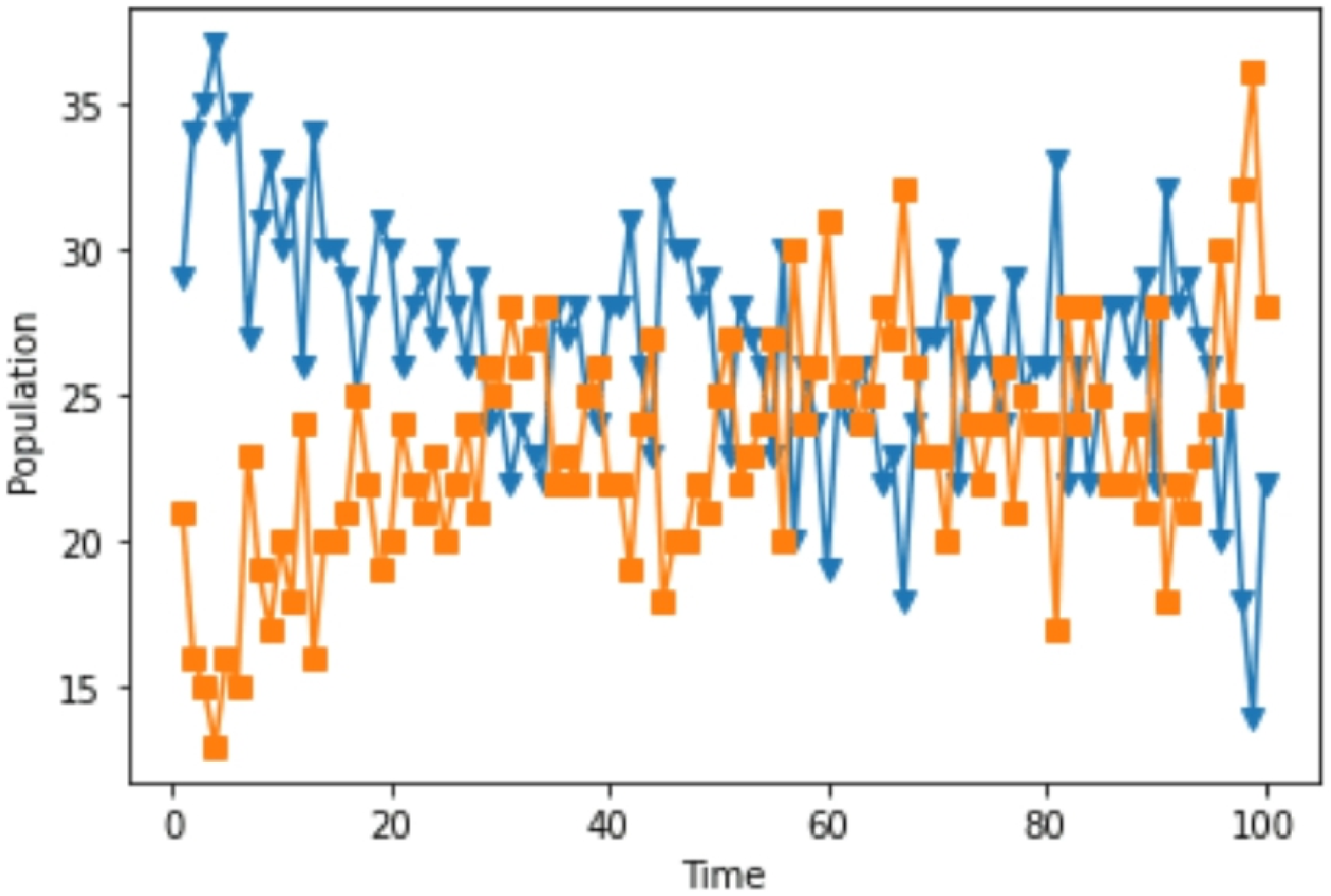}% 中括号中的为调节图片大小
	\caption{}
	\label{chutian}%文中引用该图片代号
\end{figure}
\subsubsection{Population Distribution under the opinion Fluctuation}
Since the population simulation results in the above satisfy the condition of the consensus, we set the non-convergent opinion parameters (i.e.,$d=\phi=\sigma=1$) to explore whether its distribution also has the property of "phase transition" in this part. It can be seen that when the weight of the opinion distance(i.e., 1- $\delta$ ) is small, there is no obvious change in population distribution and consensus situation; However, when the weight is large (such as 1- $\delta$ =0.7), the distribution changes dramatically: all individuals enter a specific community, while the competing community is "empty": this is due to the drastic fluctuation of opinions, which may reflect the current user strategies of some platforms. Taking Zhihu as an example, a series of highly controversial issues can be discussed through the recommendation algorithm and hot black-box operation, thus attracting a wide audience.\\

\begin{figure}[htbp]
	\centering
	\begin{subfigure}{0.325\linewidth}
		\centering
		\includegraphics[width=0.9\linewidth]{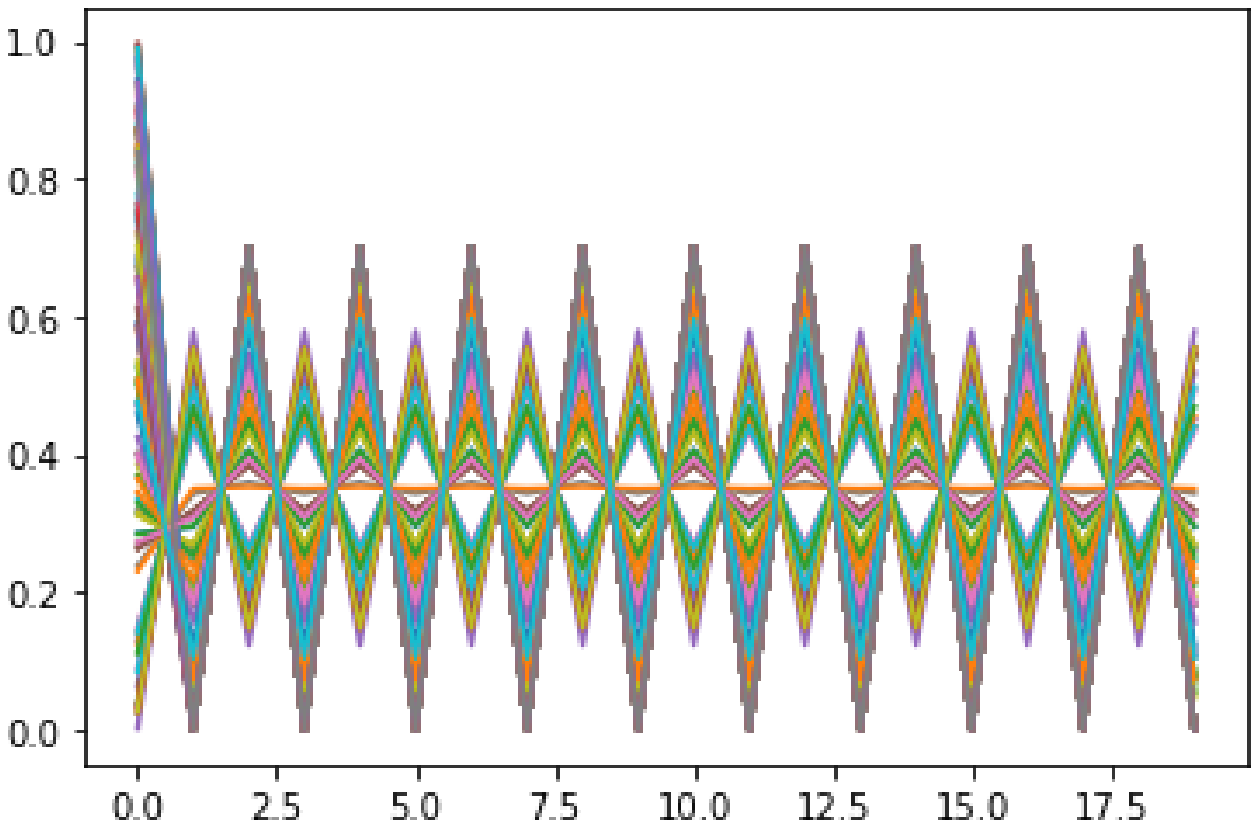}
		\caption{opinion evolution}
		\label{chutian3}%文中引用该图片代号
	\end{subfigure}
	\centering
	\begin{subfigure}{0.325\linewidth}
		\centering
		\includegraphics[width=0.9\linewidth]{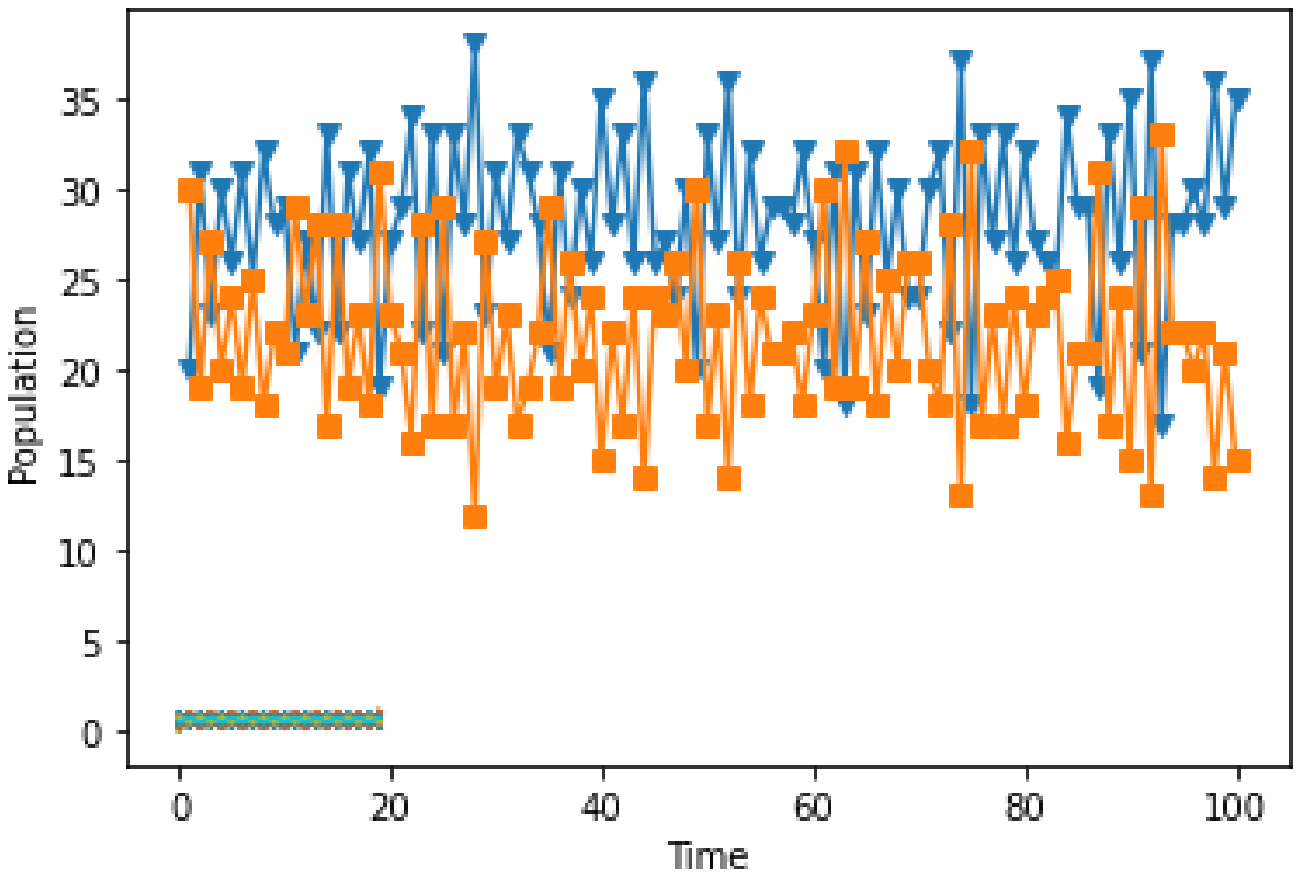}
		\caption{$\delta$= 0.8 \quad population}
		\label{chutian3}%文中引用该图片代号
	\end{subfigure}
	\centering
	\begin{subfigure}{0.325\linewidth}
		\centering
		\includegraphics[width=0.9\linewidth]{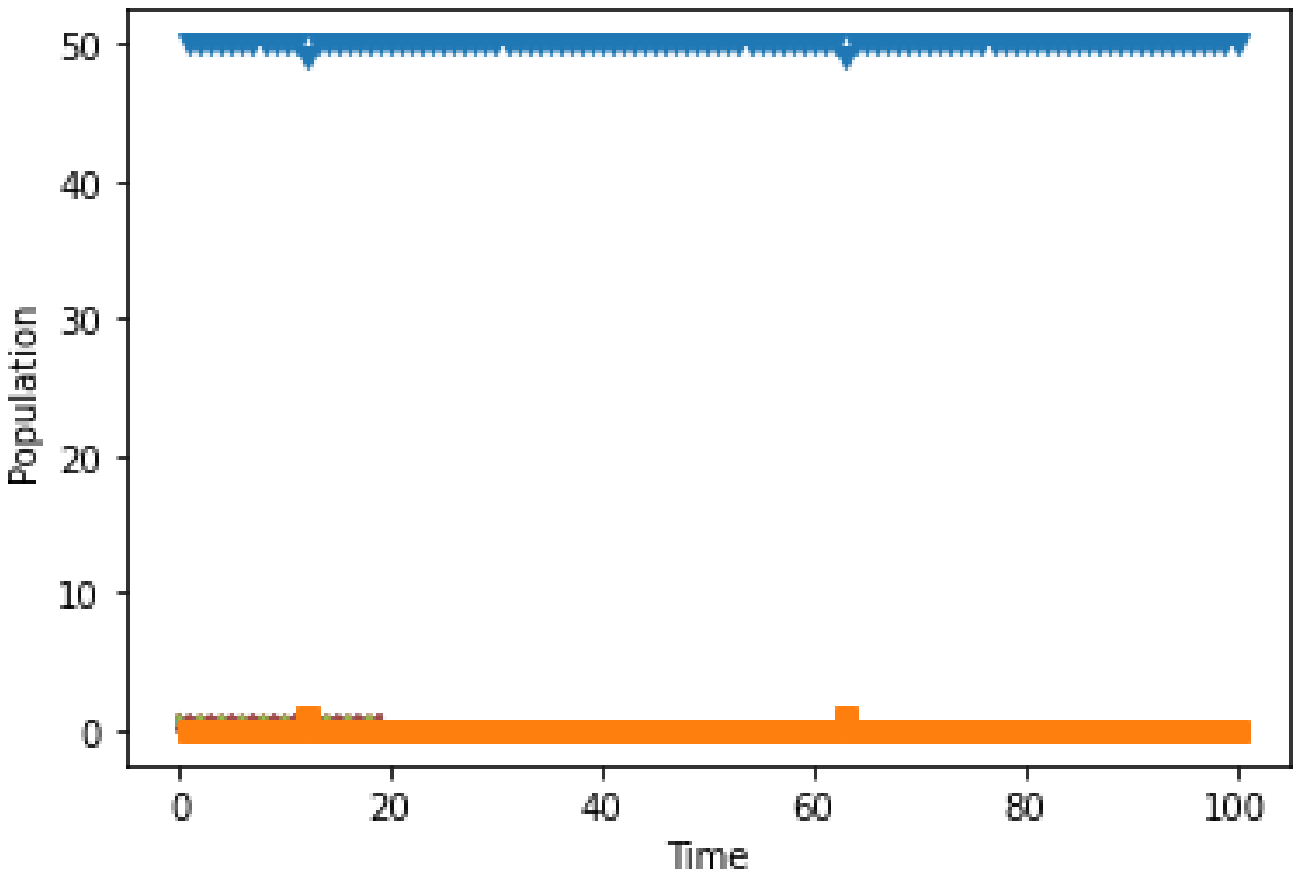}
		\caption{$\delta$= 0.3\quad population}
		\label{chutian3}%文中引用该图片代号
	\end{subfigure}
	\caption{mother of chutian}
	\label{da_chutian}
\end{figure}

\section{Conclusion and Discussion}

\subsection{Conclusion and its Expansion}

In this paper, we build a compound model of "opinion-migration" and introduce the methods of traditional demography into the study of user mobility. At the same time, we analyze the convergence of the model and prove two important conclusions: (1) When the sum of reduction rate ($\sigma$) and undisturbed degree ($\varphi$) is less than 1, individual opinions on any network will reach a quasi-consensus. (2) If the network is a star graph, the expectation of the community population will be a value only related to $\delta$. Based on this, through simulation experiments, we explore some complex conclusions, including: (1) $\sigma$ = $\varphi$ = 1 will cause sudden changes in the model, which means that opinions will fluctuate randomly in a certain range. (2) A positive correlation exists between the fluctuation range and the confidence threshold d. (3) In the small-world network, the community population does not converge, but the increase of $\delta$ will increase the migration rate. Generally speaking, differences in the network structure and opinions will have a nonlinear impact on cyber migration. However, due to the limitation of computing power, we have not simulated the large population situation, which needs to be further explored by scholars in related fields. \\

\subsection{Discussion and prospect}

From the research design of this paper, we give a mechanistic explanation for the online population mobility, that is, the influence sequence of "social distance-opinion-population," which may be applied to building a middle-level theory for user prediction and content recommendation. At the same time, as an attempt at analytical sociology, this mechanism tries to disassemble the black box between the "micro" level and "macro" level so that population distribution becomes a sudden result of the individual migration choice. In addition, there are still two problems worth discussing in our model: (1) Can we estimate or iteratively update the parameters such as $\varphi$ and $\delta$ by Bayesian modeling and other means? This is an area worth studying at the statistical level. (2) If (1) is unsuccessful or ineffective, can we use existing data mining methods to verify the assumptions and conclusions mentioned above? \\

Given the relationship between our model and social theory, we may look forward to a kind of "cyber sociology" that corresponds to population problems. In the communication behavior of the Internet, the evolution of opinions can be regarded as a micro process (i.e., the communication between n people), and various complex social properties and collective actions (such as power, social fragmentation, population, collective resistance, etc.) can be obtained by it. Through the study of the basic models, we can combine them effectively to simulate the real-world situation and prove the corresponding conclusions through data and theoretical analysis.

\bibliographystyle{IEEEtran}
\bibliography{ref}

\end{document}